\documentclass[preprints,article,accept,moreauthors,pdftex]{Definitions/mdpi} 

\usepackage{amssymb}
\usepackage{amsmath}

\def\eps{\varepsilon}

\usepackage{color}
\usepackage{soul}

\firstpage{1} 
\pubvolume{xx}
\issuenum{1}
\articlenumber{5}
\pubyear{2025}
\copyrightyear{2025}
\history{Received: October 7, 2026; Accepted: date; Published: date}

\Title{Mathematical statistics of wild mammal biomass}

\Author{Klaus M. Frahm $^{1}$\orcidA{},
Dima~L.~Shepelyansky $^{1,*}$\orcidB{}
and
José Lages $^{2}$\orcidC{}}

\AuthorNames{Klaus M. Frahm, Dima L. Shepelyansky and José Lages}

\address{%
$^{1}$ \quad Univ Toulouse, CNRS, Laboratoire de Physique Th\'eorique,  
Toulouse, France\\
$^{2}$ \quad Université Marie et Louis Pasteur, CNRS, Institut UTINAM (UMR 6213),
équipe de physique théorique, Besançon, France}

\corres{Correspondence: dima@irsamc.ups-tlse.fr; Tel. +33-56155-60-68 (D.L.S.)}

\abstract{Using the recently published global census of the biomass of wild
terrestrial mammals, we perform a detailed mathematical statistical
analysis of its distribution over $N_s=4795$ species, drawing on tools
developed in economics to characterize wealth inequality. We show that
the Lorenz curve of the mammal biomass distribution is 
characterized by a large Gini coefficient $G=0.944$ exceeding the 
inequality reported for wealth distribution among world countries.
This distribution is compared to the predictions of the Wealth
Thermalization Hypothesis (WTH), in which species biomass values are
treated as energy levels populated according to a Rayleigh-Jeans (RJ)
steady-state distribution. We show that an RJ extended spectral model
reproduces the real Lorenz and Pareto curves over four orders of
magnitude of biomass variation, capturing the strong condensation of
biomass among the rare, heavy-bodied species and its near-absence
among the vast majority of light-bodied ones. These results
extend the WTH framework, previously validated on distributions of
human economic origin, to a biological distribution shaped by
ecological and evolutionary constraints, and place the extreme rarity
of large-bodied mammal species within the same statistical mathematical
description as the oligarchic concentration of wealth in human
societies. We finally discuss possible ecological mechanisms
that may contribute to the observed distribution,
including interspecific interactions, food availability, and competition.}


\begin{document}
	
\section{Introduction}
\label{sec1}
	
Recently, an exhaustive analysis of the biomass of wild terrestrial
mammals was reported in \cite{milopnas,milogit}.
For the land wild mammals the biomass data is available for $N_s=4795$ different species.
The biomass of marine mammals is also presented in this work
but this data is not so detailed due to a more difficult process of its determination.
While the global properties of the biomass distribution
are discussed in \cite{milopnas,milogit} it is useful to perform  a more detailed
mathematical statistical analysis  of this distribution.

The strong inequality between rare, heavy-bodied species and abundant,
light-bodied ones is a long-standing theme in ecology. Macroecological
studies have shown that population density scales inversely with body
mass across mammal species \cite{damuth}, a pattern related to the
allometric scaling of metabolic rate with body size \cite{kleiber}:
fewer individuals of a large-bodied species can be sustained by a
given amount of resources than individuals of a small-bodied one.
This mechanism, together with the higher vulnerability of large
mammals to hunting, habitat loss and slow reproduction, has driven a
severe decline of the world's largest herbivore species in recent
decades \cite{ripple}. This well-documented ecological
asymmetry between the few heavy species
and the many light ones calls for a compact quantitative statistical
description, for which we turn to tools developed in a seemingly
distant field: economics.

For the detailed analysis of this statistical biomass distribution, 
we apply tools broadly used in economic studies
of the wealth inequality in world countries and
the whole world \cite{piketty1,piketty2,boston,yakovenko2}.
In this scientific field one typically uses the Lorenz curve to 
characterize a given wealth distribution. This curve 
\cite{lorenz,boston} represents the dependence of cumulated 
normalized wealth $0\leq w \leq 1$ on the cumulated normalized fraction of
population or households $0 \leq h \leq 1$.
The  perfect equipartition of wealth corresponds to
the diagonal $w=h$ and the doubled area between diagonal
and the Lorenz curve $w(h)$ gives the Gini coefficient
$0 \leq G \leq 1$  \cite{gini,boston}.

Values of Gini coefficients $G$ for world countries in 2021 
are presented in \cite{wikigini} being in the range
$0.59 < G < 0.90$ while for the whole world $G = 0.889$.
High $G$ values  close to unity correspond to
a high inequality in a country with a large fraction of poor population
and a tiny oligarchic population fraction that owns
a huge fraction of the total wealth \cite{piketty1,piketty2}.
The distribution of high revenues
is also described by the Pareto distribution curve \cite{pareto,yakovenko2}.

The astonishing feature for the wealth distribution of world countries
is the strong inequality of wealth of households
as reviewed e.g. in \cite{piketty1,piketty2,boston}.
Thus, according to \cite{piketty2}
for the whole world 
50\% of the population owns only 2\% of total wealth, while the top 10\%
owns 75\% of total wealth, of which 38\% is held by the top 1\% alone.

In this work, we trace an analogy between the wealth of households
and the biomass of wild species, and apply the Lorenz curve
description to the biomass data available at \cite{milopnas,milogit}.
As shown below, the resulting biomass distribution has a very high Gini
coefficient $G=0.944$, comparable to the most unequal wealth
distributions found among world countries.

Recently, the Wealth Thermalization Hypothesis (WTH)
was proposed in \cite{jsp} with the aim to explain the origins of
such universal wealth inequality existing in countries
and the whole world. The WTH argues that various types of links and nonlinear 
interactions between society members (agents) in a given country
lead to a dynamical thermalization over layers of wealth
corresponding to a social stratification in a society \cite{stratif}.
Within the WTH framework, the wealth values $w_m$ can be considered as
energies $E_m$ of states $m$ (or related frequencies).
The thermalization appears from the classical dynamical
nonlinear equations leading to chaos. The process  
has two conserved quantities being total system
energy $E = \sum_m E_m \rho_m$ and probability norm 
$\eta = \sum_m \rho_m =1$ (rescaled number of agents)
where $\rho_m$ are the time averaged squared amplitudes and occupation 
probabilities of each oscillator mode $m$ \cite{jsp}.  
According to the well known results
of thermodynamics of classical fields \cite{landau,zakharovbook} 
the steady-state distribution of probabilities (or fractions)
for the Rayleigh-Jeans (RJ) thermalization has the form:
\begin{equation}
	\rho_m = \frac{T}{E_m-\mu} \qquad ({\rm RJ}) .
	\label{eqrj}
\end{equation}
Here $\rho_m$ represents averaged stationary 
probabilities at certain wealth states 
$0 \leq m < N$ with wealth/energies $w_m=E_m$. The 
two parameters correspond to the system temperature $T$ and 
the chemical potential $\mu(T)$ which are 
determined by the values of the total energy $E$ and the norm $\eta=1$
via both equations given above (see e.g. Refs. 
\cite{jsp,rmtprl,ourfiber,complex} for more details).
Within this analogy between the biomass $b_m$ of a species and
state wealth $w_m$ or energies, 
we assume equivalence $b_m\leftrightarrow w_m=E_m$.
In the following, we mainly use $w_m$ for notations of biomass values $b_m$.

The surprising feature of the thermal RJ distribution (\ref{eqrj})
is the appearance of RJ condensation when at low values of total energy $E$
a huge amount of total probability is condensed at the low energy states.
The condensation fraction can be very close to unit value $\eta=1$
(see e.g. \cite{jsp,ourfiber,complex}).
This condensation provides a natural explanation of the emergence of 
poor and oligarchic phases in the wealth inequality \cite{jsp,complex}.
The existence of RJ condensation was first pointed out by 
H.~Fr\"ohlich in 1968 for electric modes in biocells 
\cite{frohlich1,frohlich2}
(even if there is a logical gap in these works as pointed out 
in \cite{jsp,ourfiber,complex}).
For multimode optical fibers, RJ condensation
was found in numerical simulations \cite{picozzi1} and experimentally observed in
\cite{wabnitz,picozzi2}.
We also point out that the RJ condensation belongs to
a more generic phenomenon of statistical mechanics known as the
constraint-driven condensation \cite{trizac,satya,marsili} which describes
such diverse physical systems as
coalescence in granular media, jamming in traffic, 
gelation in networks \cite{satya}
and financial data analysis \cite{marsili}.

In works \cite{jsp,complex} it was shown that the WTH approach
based on the RJ thermalization and condensation provides
a good description of wealth inequality distributions in countries, of
Gross Domestic Product (GDP) across world countries, of market
capitalization of companies of the Hong Kong, Shanghai and London stock
exchanges, and of bitcoin transactions and world trade flows.
It is also shown that the RJ thermalization and condensation (\ref{eqrj})
well describe the energy and carbon emission distribution over countries (arXiv:2506.17720 [cond-mat.stat-mech]),
distribution of votes in EU elections (arXiv:2506.17720 [cond-mat.stat-mech]),
and even distribution of given names in USA and France (arXiv:2508.06048 [cond-mat.stat-mech]).
In this work, using the data of \cite{milopnas,milogit}, 
we show that the WTH concept also provides a good description
of the biomass distribution over species of wild mammals.

The paper is composed as follows: Section~\ref{sec2} presents the
construction of the Lorenz and Pareto curves from the biomass data and
recalls the details of WTH approach; Section~\ref{sec3}
presents the resulting comparison between real data and the WTH models
for land wild mammals; Section~\ref{sec4} discusses the results
and their interpretation.
	
\section{Construction of Lorenz and Pareto curves and WTH theory for the mammal biomass}
\label{sec2}

We first describe, the construction of the Lorenz curve from the 
{\em real biomass data} \cite{milopnas,milogit} and give afterwards 
the modifications to obtain Lorenz curves from the theoretical RJ models 
(see below). 
Let us consider the ordered set of normalized biomasses 
(\textit{spectrum of biomass}),
$w_0<w_1<\ldots<w_m<\ldots <w_{N_s-1}$ to which we attribute 
uniform probabilities $\bar\rho_m=1/N_s$, $\forall m$,
where $N_s=4795$ is the total wild mammals species \cite{milopnas,milogit}. 
Otherwise stated, $w_m=b_m$ is the biomass of the 
species $m$ and $\bar\rho_m$ its proportion or household fraction. 
We construct the (real data) Lorenz curve 
by computing the {\em cumulated normalized} 
household fraction of species as $h(m) =\sum^{m-1}_{l=0} \bar\rho_l$
and the {\em cumulated normalized} biomass/wealth fraction 
$w(m) =(1/w_s) \sum^{m-1}_{l=0} w_l \bar\rho_l$. Here $w_s$ is the 
average mammal biomass given by 
$w_s = \sum_{m=0}^{N_s-1} w_m\bar\rho_m=(\sum_{m=0}^{N_s-1} w_m)/N_s$ 
and we use the notations $h(m)$ and $w(m)$ for the {\em cumulated} 
and {\em normalized} household and wealth fractions while $w_m$ designate the 
biomass of the specific species $m$. 
This procedure provides the {\em real data} Lorenz curve with $N_s+1$ 
data points $(h(m),w(m))$ (for $m=0,1,\dots,N_s$) 
such that $h(0)=w(0)=0$ and $h(N_s)=w(N_s)=1$.


Now, we consider a theoretical model spectrum of $N$ oscillator energies $E_k$
(with $0=E_0<E_1<\ldots<E_{N-1}$) representing $N$ agents. We assume that 
this system is thermalized at the global energy $E=\sum_k E_k\rho_k$ 
and with the conserved norm $\eta=1=\sum_k \rho_k$. Here 
$\rho_k$ is given by the RJ expression $\rho_k=T/(E_k-\mu)$ 
where $T$ and $\mu$ are determined by the two implicit equations 
for the energy and the norm (see e.g. \cite{complex} for details on this). 
For this spectrum, we can construct a {\em theoretical RJ} Lorenz curve in a 
similar way as above by $h(k)=\sum_{l=0}^{k-1} \rho_l$ and 
$w(k)=\sum_{l=0}^{k-1} (E_l/E)\rho_l$. However, here we use the given 
model energies $E_k$ and the non-uniform RJ probabilities 
$\rho_k=T/(E_k-\mu)$. 
This construction procedure is invariant with respect to a simple energy 
rescaling $E_l\to\alpha E_l$, $E\to \alpha E$ (and similarly for 
$T$ and $\mu$). Therefore, the relevant 
parameter is the rescaled energy $\eps=E/B$ where $B=E_{N-1}$ is 
energy band width. We also fix $E_0=0$ which is reasonable to modelize 
the biomass distribution since $w_0/w_{N-1}\sim 10^{-11}$. 
To be more precise, we assume that 
multiple agents (with index $k$) describe real values $\bar\rho_m, w_m$  
for a given mammal (with index $m$). 

Now, we specify two specific models for the precise form of the 
model energy spectrum. The simplest assumption is that the density of 
energy states
$\nu(k) =d k/d E_k$ is constant so that
$0 \leq E_k=k/N < B=1$ for $k=0,1,2,\ldots,N-1$. This model
is called the RJ standard (RJS) model \cite{ourfiber,jsp}. 
For sufficiently small values of $\eps=E/B$, it shows RJ condensation 
where a significant fraction of probability is concentrated on a 
small number of modes at minimal $k$ values. It provides Lorenz curves 
that agree on a qualitative level with  Lorenz curves from real data. 

To obtain a more precise description, we use a more advanced
model called the RJ extended (RJE) model with 
$E_k =(e^{ak/N}-1)/(e^a-1)$ where $a$ is 
a real parameter. The limit $a\to 0$ gives the RJS model 
with $E_k=k/N$ and for $a>0$ the density of states is decreasing at 
high energies according to $\nu(E_k)=dk/dE_k = N(e^a-1)/[a(1+(e^a-1)E_k)]$; 
the global energy scale here is chosen such that $B=E_{N-1}\approx 1$. 
For the wealth inequality the decay of $\nu$ at high energies is 
rather natural since there are fewer rich people (or countries) at high 
wealth/energy values.
We consider that this is also natural for the biomass distribution
where the species with high biomass are more rare. 
In this work, we mainly present results for the RJE model.
	
To recover the Lorenz curve from the RJ theory (\ref{eqrj}), 
one can easily apply the above procedure for a spectrum with large 
but finite $N$, e.g. $N=10000$ or larger values. 
It is also possible to obtain analytical formula for 
the continuous limit $N\to\infty$ in terms of the chemical potential 
$\mu$ which is determined by an implicit equation as a function of 
$\eps$ and $a$ and the Lorenz curves are very stable for sufficiently 
large values of $N$ (and agree with the continuous limit $N\to\infty$)
at given parameters $a$ and $\eps$ (see details in \cite{complex}). 
The RJS/RJE Lorenz curves (and other related RJ quantities such as $G$ 
and the Pareto curves; see below) presented  
in this work, were computed from the continuous limit $N\to\infty$.
	
To compare the RJS and RJE models with real Lorenz curves, we choose 
the parameter $\eps$ such that the RJ and real data Lorenz curves have the 
same Gini coefficient $G$. For the RJE model, we also minimize the geometrical 
curve distance between both curves to determine the optimal 
choice for the parameter $a$ 
(under the constraint that for each value of $a$ the parameter $\eps$ 
is recomputed to provide the same Gini coefficient $G$ as the real data). 
This procedure is the same as the one used in \cite{jsp,complex}. 

In addition to the Lorenz curve, it is also useful to introduce  
another quantity which stresses more the details for the very 
rich population or the heaviest biomass population. This 
quantity is the cumulative distribution function (CDF)
$C(w_m)$, also called {\em Pareto distribution} (in this work), 
that gives the fraction of households 
(species, companies, people, countries etc.) having a wealth (``biomass'') 
larger than $w_m$.
This quantity is typically used in economics 
(see e.g. \cite{pareto,yakovenko2}) 
and it can be directly obtained 
by drawing $C(w_m)=1-h(m)$ versus $w_m$ (for both cases 
of real data and RJ models). Usually the notation Pareto distribution 
is applied to the special algebraic form $C(w_m)=(w_0/w_m)^{-\alpha}$ 
with  two parameters $w_0>0$ and, typically, $\alpha>1$. In our work, 
we simply use the notation 
{\em Pareto distribution (or curve)}  for the CDF $C(w_m)$, 
for other more general cases even if there is no 
simple power law. Thus we use the Lorenz and Pareto curves
to describe the properties of biomass distribution.

In contrast to the Lorenz curve, the Pareto curve 
is not scale independent and in order to compare real data with 
possible RJE curves, it is necessary to fix the global energy/biomass 
scale either by the maximal or the average value of $w_m$. Here, we show 
$C(w_m)$ as a function of $w_m/\langle w_m\rangle$ where for the 
RJS/RJE models, we simply have $\langle w_m\rangle=\eps$ (since $B=1$ and 
$E_m=w_m$). For real data the average value $\langle w_m\rangle$ 
is more reliable since the 
maximal value of $w_m$ for a given data set may be subject to 
strong statistical fluctuations (note that 
$\max(w_m)=1$ for the RJS/RJE models 
since $B=1$). 

\begin{figure}[t]
\begin{center}
\includegraphics[width=1\textwidth]{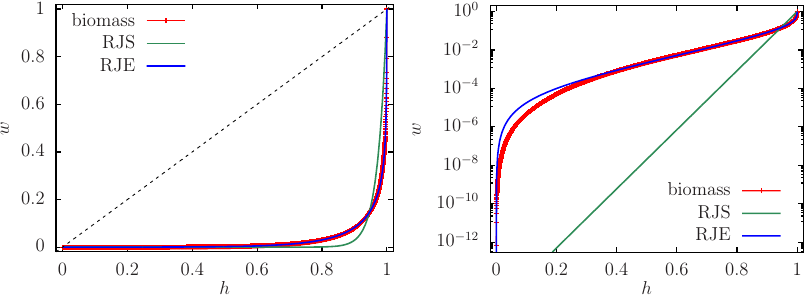}
\end{center}
\caption{\label{fig1}
Comparison of mammal biomass Lorenz curve (red data curve) with 
the theoretical RJS model (green curve) and the RJE model at optimal parameter 
$a=7.034$ (blue curve). The $x$-axis corresponds to the 
cumulated fraction of species ($h$) and the $y$-axis to
the cumulated fraction of biomass ($w$). 
The dashed black line (only in left panel) 
corresponds to the line of perfect equipartition $w=h$. 
For both cases of RJS and RJE, the corresponding 
values of $\eps$ are determined such that the theoretical curves have the 
same Gini coefficient $G=0.944$ as the red data curve. This gives 
$\eps_{\rm RJS}=0.0282$ and 
$\eps_{\rm RJE}=7.57\times 10^{-4}$. The left panel shows the Lorenz curves 
in normal representation and the right panel in logarithmic representation 
for the biomass variable $w$. Note that the red top data points for 
$h\approx 1$ visible in the left panel correspond to the species with 
largest biomass while the lowest data points for $h\approx 0$ 
in the right panel correspond to the species with minimal biomass 
(see also discussion in text about top/bottom biomass species). 
The green RJS curve (straight line in the right panel) can be very accurately 
approximated by the simple exponential expression 
$w_{\rm RJS}(h)\approx e^{(h-1)/\eps_{\rm RJS}}$ which was obtained for 
values $\eps_{\rm RJS}<0.2$ in \cite{complex}.}
\end{figure}
	
\section{Results}
\label{sec3}
	
The publicly available biomass data of \cite{milopnas,milogit} 
concern $N_s=4795$ wild land mammals with a total biomass of 
$b_{\rm tot}=2.24\times 10^{13}$g. 
The species with the 3 heaviest biomasses are:
White-tailed deer
({\it Odocoileus virginianus}) with biomass $b_1 \approx 2.68\times10^{12}\,$g;
Wild boar ({\it Sus scrofa}) with $b_2 \approx 1.95\times10^{12}\,$g; and
African bush elephant ({\it Loxodonta africana})
with $b_3 \approx 1.33\times10^{12}\,$g 
corresponding to 12\%, 8.7\% and 5.9\% respectively of the total biomass. 
The three lightest biomasses are:
New Caledonian long-eared bat ({\it Nyctophilus nebulosus}) with $b_{4795} = 14.4 g$;
Madeira pipistrelle ({\it Pipistrellus maderensis}) with $b_{4794} = 218.9 g$; and
Sri Lankan rain forest shrew ({\it Crocidura hikmiya}) with $b_{4793} = 466.1 g$.
Thus the biomass variation is exceptionally large
changing by an approximate factor $2 \times 10^{11}$. 
This variation is much larger than all
cases studied previously in \cite{jsp,complex}. 
As a result the values $w$ of the Lorenz curve are changing from 
$w(1)=6.43\times 10^{-13}$ (at $h(1)=1/N_s=2.085\times 10^{-4}$) to $w(N_s)=1$ 
(at $h(N_s)=1$).
	
Fig.~\ref{fig1} compares the RJS and (optimal) RJE Lorenz curves with 
the real Lorenz curve. 
The RJS model captures the existence of the RJ condensation phase
but it still has significant deviations from the real data. 
On the other hand, the RJE model, at optimal choice for the parameter $a$, 
describes very well the real data 
for biomass cumulated value in the range $10^{-4} \leq w \leq 1$.
There are visible deviations for $w < 10^{-4}$ that we attribute to
weak thermalization for species with small biomass.
Also some species can be located only in specific isolated places on Earth 
where there are only weak interactions between species
as it is the case e.g. for the Madeira pipistrelle 
({\it Pipistrellus maderensis}).
	
\begin{figure}[t]
\begin{center}
\includegraphics[width=\textwidth]{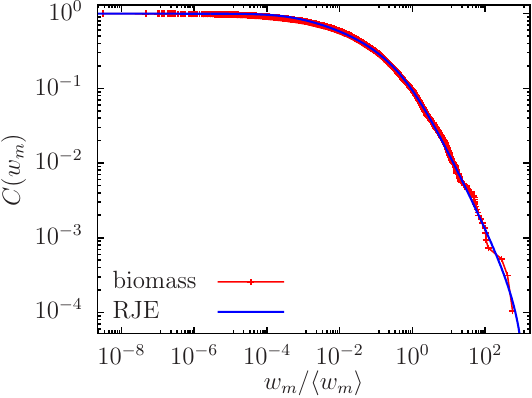}
\end{center}
\caption{\label{fig2}
Pareto curves $C(w_m)$ for the same data of Fig.~\ref{fig1} (red 
data curve) where  $C(w_m)$ represents the fraction of mammal species with a biomass 
larger than $w_m$ (analogous to wealth). 
The $x$-axis corresponds to rescaled 
values $w_m/\langle w_m\rangle$ where 
$\langle w_m\rangle = w_s=4.68\times 10^{9}$g is the 
average species biomass. 
The blue curve corresponds to the 
theoretical RJE Pareto curve with same parameters as in Fig.~\ref{fig1}
($a=7.034$ and $\eps_{\rm RJE}=7.57\times 10^{-4}=\langle w_m\rangle$). 
Note that for the theoretical RJE model, we have $w_m\in[0,1]$ 
(see text and \cite{complex} for details).}
\end{figure}
	
Fig.~\ref{fig2} compares the Pareto curves of the RJE model 
with the real data curve. 
It shows that the WTH approach describes very well the 
real biomass data in the range of 4 (10) orders of magnitude
of variation of $C(w_m)$ ($w_m$). 
	
\begin{figure}[t]
\begin{center}
\includegraphics[width=\textwidth]{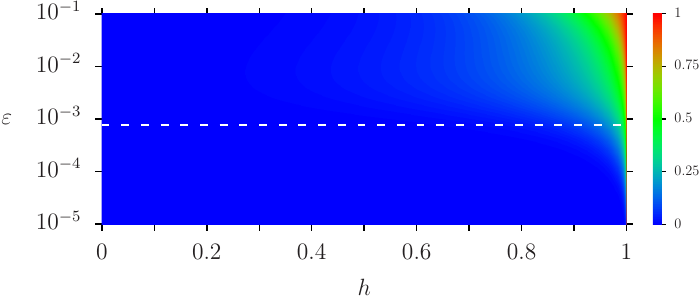}
\end{center}
\caption{\label{fig3}
Color plot of cumulated wealth $w$ from the Lorenz curves of the RJE model 
for the parameter $a=7.034$. The $x$-axis corresponds to 
the fraction of households $h\in[0, 1]$ 
and the $y$-axis to the rescaled 
energy $\varepsilon=E/B\in[10^{-5}, 10^{-1}]$ in logarithmic representation. 
The values of the colorbar represent the cumulated wealth fraction 
$w\in[0,1]$ at given values of $h$ and $\eps$. 
The white dashed line corresponds to the value 
$\eps_{\rm RJE}=7.57\times 10^{-4}$ used in Figs.~\ref{fig1} and \ref{fig2} 
for the blue RJE curves (with the same value of $a$). 
Note that for $a=7.034$ the critical energy value 
for the transition from positive to negative temperature is given 
by $\eps_C=0.141$ (slightly above $0.1$) and that 
the color values along the dashed line correspond to the blue RJE 
curve in Fig.~\ref{fig1}.}
\end{figure}

The global dependence of the cumulated biomass $w$
on the cumulated number of species $h$ and the average rescaled
total system energy $\eps$
is shown by color in Fig.~\ref{fig3} at fixed $a=7.034$. 
The white line in this Fig.~\ref{fig3}
shows the parameter value $\eps=7.57 \times 10^{-4}$
corresponding to the real biomass data. 
This Figure shows the existence of the huge phase of very light
species corresponding to the phase of poor households
in the problem of wealth inequality.
Thus the bottom 50\% of wild mammals species 
have only $0.23$\% of total biomass while the top 
10\% (1\%; 0.1\%) of species capture $92$\% ($63$\%; $31$\%) of the 
total biomass. (These numbers refer to the real data but the RJE 
Lorenz curve at optimal value of $a$ gives very close values.)
These values show an even stronger biomass inequality
in the world of wild mammals as compared to 
the case of wealth inequality in human society \cite{piketty1,piketty2,boston}.
	
The biomass of marine mammals is hardly accessible and we do not discuss
the marine biomass distribution here.
The data  in Table~2 of \cite{milopnas}
is given only for the 10 heaviest marine biomass values
and there are no reliable data for a great majority of marine species.

\section{Discussion}
\label{sec4}

In this work, we presented a mathematical statistical description
of biomass distribution for
wild mammals using the public data reported in \cite{milopnas,milogit}.
We show that this distribution
can be well characterized by
Lorenz and Pareto curves, tools broadly used
in the economic studies of wealth inequality \cite{piketty1,piketty2,boston}. 
The Lorenz curve of land mammal biomass has a very high Gini coefficient 
$G=0.944$ 
corresponding to a very high inequality of biomass distribution
between species.

We also show that the WTH approach introduced in \cite{jsp,complex}
gives a good description of the biomass distribution.
This extends a 
body of previous results in which the same WTH description was shown
to apply to a wide range of systems related to human activity
\cite{jsp,complex} -- from wealth and GDP
distributions to elections and given names -- and we show here that
it works equally well for the biomass of wild species, where the
influence of humans remains rather restrictive.

This recurrence across such diverse systems supports the view that
the precise mechanism of thermalization is not crucial for the
emergence of the RJ steady-state distribution (\ref{eqrj}). The
mechanisms at play between species can themselves be diverse,
ranging from food-chain relations, in which one species
preys upon another, to competition for pastures and other
shared resources.
A close analogy is found in statistical physics: for
an atomic gas in a
three-dimensional box, almost any weak interaction between atoms
drives the system to the Boltzmann-Gibbs distribution of kinetic
energies, with an average value $3kT/2$ where $k$ is the Boltzmann
constant, largely independently of the microscopic details of the
interatomic potential. In the same spirit, we expect the RJ
thermalization of biomass to be a robust, interaction-independent
outcome of the network of trophic and competitive links between
species, rather than a fine-tuned consequence of any particular
ecological mechanism.

From an ecological viewpoint, the strong concentration of biomass in a small number of heavy-bodied species,
together with the low biomass associated with the large number of light-bodied species, is consistent with
well-established macroecological relationships between body mass and population density \cite{damuth,kleiber}.
Large-bodied mammals generally occur at lower population densities, while the largest species are also particularly
vulnerable to population declines and extinction \cite{ripple}. Rather than singling out any one of
these as the driver of thermalization -- consistent with the mechanism-independence argued above -- we note
that the resulting RJ condensation, within the WTH framework, is compatible with these independently
documented macroecological regularities. The species forming the high-biomass tail of the distribution
are also those for which ecological and anthropogenic pressures may have particularly strong consequences.

	\authorcontributions{All authors equally contributed to all stages of this work.
	}
	
	\funding{The authors acknowledge support from 
		NANOX $N^\circ$ ANR-17-EURE-0009 in the framework of 
		the Programme Investissements d'Avenir (project MTDINA).
	}
	

	\conflictsofinterest{The authors declare no conflict of interests.
	} 
	


		\reftitle{References}

	\end{document}